\documentclass[twocolumn]{aastex631}

\newcommand{\lum}{${\rm~ergs~s^{-1}}$}

\def\gtrsim{\lower 2pt \hbox{$\, \buildrel {\scriptstyle >}\over
{\scriptstyle \sim}\,$}}
\def\lesssim{\lower 2pt \hbox{$\, \buildrel {\scriptstyle <}\over
{\scriptstyle \sim}\,$}}

\def\hst{{\sl HST}}

\def\chandra{{\sl Chandra}}

\def\hst{{\sl HST}}

\newcommand{\as}{$^{\prime\prime}$}

\def\approxlt{\lower.2em\hbox{$\buildrel < \over \sim$}}
\def\approxgt{\lower.2em\hbox{$\buildrel > \over \sim$}}

\def \ms   {\hbox{M$_{\odot}$}}
\def \jwst{\hbox{\it JWST}}

\def \xs{\hbox{X-1}}
\def \sou{\hbox{KUG 1138+327}}
\newcommand\z{$z$}

\begin{document}

\title{Candidate intermediate-mass black hole discovered in an extremely young low-metallicity cluster in the tadpole galaxy KUG 1138+327}

\author[0000-0002-9279-4041]{Q. Daniel Wang} \thanks{Contact e-mail:wqd@umass.edu} 
\affiliation{Department of Astronomy, University of Massachusetts, Amherst, MA 01003, USA}
\affiliation{Institute of Astronomy and Astrophysics, Academia Sinica, No.1, Sec. 4, Roosevelt Road, Taipei 10617, Taiwan, R.O.C.}
\author{Juergen Ott}
\affiliation{National Radio Astronomy Observatory. 1011 Lopezville Road. Socorro, NM 87801, USA}

\begin{abstract}
We explore what unusual products a starburst of about 6\% solar metallicity and a mean estimated age of $\sim5 \times 10^5$~yrs  can produce in KUG 1138 + 327 at a distance of 24.5 Mpc. Chandra X-ray observations show a dominant point-like source with an average 0.3-10 keV luminosity of  $10^{40.3}$\lum\ and variability by a factor of $\sim 2$ over months. This extreme ultraluminous X-ray source (ULX) is apparently associated with the young central cluster. A multicolor disk modeling of the X-ray spectrum of the source suggests a standard accretion around a black hole. It also has a morphologically elongated nonthermal radio continuum counterpart on the scale of $\sim 200$~pc, probably the longest detected from such a source. The radio, optical, and X-ray findings suggest that it could well be an intermediate-mass black hole undergoing sub-Eddington accretion from a massive star companion. Accounting for the presence of the ULX and the prominent emission lines He\texttt{II}$\lambda$4658 and [Ar\texttt{IV}]$\lambda$4711 while lacking Wolf-Rayet spectral features, we estimate the true age of the starburst to be about 2-4 Myr. Only with such a moderate age can the starburst host this extraordinary ULX, probably triggered by a recent influx of extremely low-metallicity gas. This study demonstrates the potential of multiwavelength studies of low-metallicity starbursts to provide insights into what may commonly occur in high-redshift galaxies. 

\end{abstract}


\section{Introduction} \label{s:intro}

Metallicity affects the formation and evolution of massive stars and their end products \citep[e.g.,][]{Belczynski2010,Linden2010,Mapelli2013,Ou2023,Burrows2024,Deng2024}. 
Metals have strong effects on the opacity, dust content, and cooling rate of gas clouds, and hence the formation efficiency of stars and potentially their initial mass function \citep[e.g.,][]{Marks2012}. At lower metallicities, massive stars produce weaker stellar winds and form more massive black holes (BHs). Binaries can be more compact, forming more high-accretion-rate Roche lobe-overflow systems. Higher masses of stars or their remnants in a stellar cluster further shorten the time scales of dynamic exchange, and hence accelerate the BH binary formation and evolution.  The net effect is not only the production of more high-mass X-ray binaries (HMXBs), but also the presence of more massive and hence more luminous ones. Such low-metallicity HMXBs may be responsible for the formation of black hole binaries of masses $\gtrsim 30-60$ \ms, detected via gravitational wave emission from their coalescence \citep[e.g.][]{Belczynski2020,Abbott2020}. In the most extreme cases, low metallicity can lead to the formation of the so-called intermediate-mass black holes (IMBHs) in the mass range of $\sim 10^2  - 10^6  M_\odot$, either via direct collapse or a dynamical process \citep[e.g.,][]{Askar2023,Chiaki2023,Rantala2024,Vergara2024}.

Particularly interesting HMXBs are the so-called ultraluminous X-ray sources (ULXs), which are defined to have the 0.3-10 keV luminosity of $L_X > 10^{39} {\rm~erg~s^{-1}}$,  exceeding the Eddington limit for typical stellar-mass compact objects (neutron stars and BHs of $\lesssim 10~M_{\odot}$), under the assumption of isotropic emission \citep[e.g.,][]{Fabbiano1989, Kaaret2017,King2023}. The high luminosity of ULXs is usually hypothesized
to originate from massive accreting objects \citep[e.g. IMBHs;][]{Colbert1999,Wang2002} or super-Eddington accretion, as strongly suggested by the discovery of
pulsation \citep{Bachetti2014} and ultra-fast ($\sim 0.2$c) outflows \citep{Pinto2016}. The presence of IMBHs 
is first confirmed in gravitational wave events \citep{Abbott2020}. Such BHs are suggested to be the best explanations for the brightest ULXs or hyperluminous X-ray sources (HLXs) with $L_X \gtrsim 10^{41} {\rm~erg~s^{-1}}$ \citep[e.g.,][]{Gao2003,Pasham2014,Mezcua2015,Barrows2024} and possibly many less bright, but nevertheless extreme ULX with $L_X$ in the range of $10^{40} - 10^{41} {\rm~erg~s^{-1}}$ \citep[e.g.,][]{Roberts2023}.  Therefore, ULXs allow us to probe extreme accretors and accretion processes, which are still poorly understood. 

\begin{figure*}[!htb]
\centerline{
\includegraphics[width=1\linewidth,angle=0]{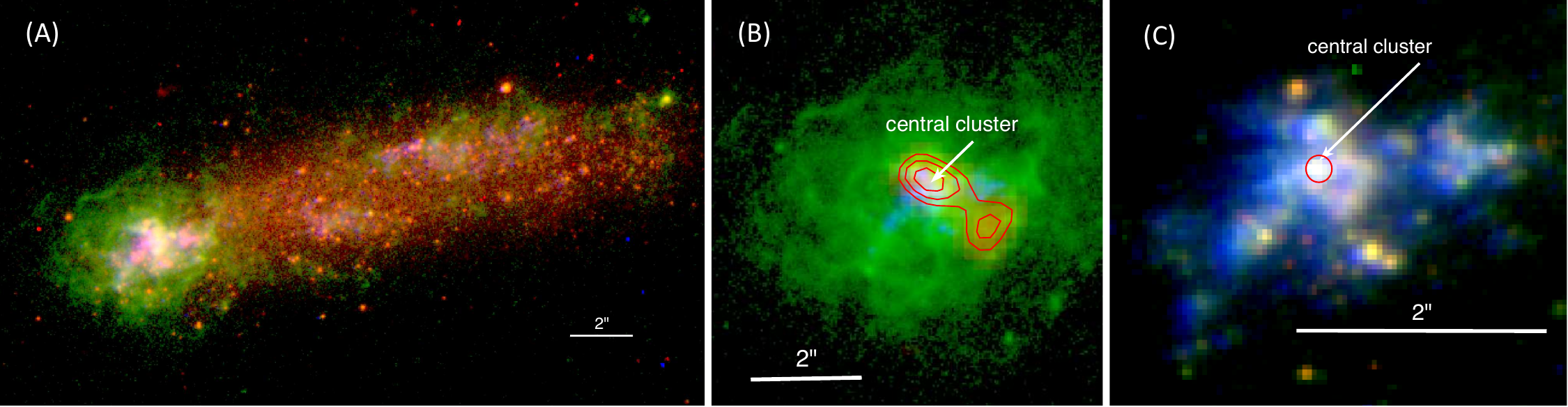}}
\caption{
3-color imaging views of \sou: {\bf (A)} \hst\ WFC3 F814W (red), F657N (green) and F336W (blue); north is up, east to the left. {\bf (B)} Close-up of the starburst head: VLA A-array 1.5 GHz  (red), \hst\ F657N (green) and F225W (blue). The radio intensity contours are at (6, 7, and 8) $\times 10^{-5} {\rm Jy/beam}$.{\bf (C)} Close-up of the stellar cluster emission in the starburst head: \hst\ WFC3 F814W (red), F547M (green) and F225W (blue). The red circle marks the position of our detected source \xs, consistent with that of the central (white) stellar cluster, which is the youngest and most massive in the galaxy \citep{Elmegreen2016}. 
 }
\label{f:f1}
\end{figure*}

At a distance of 24.5 Mpc (1\as\ = 119~pc), \sou\ (also known as LEDA-36252 and kiso\,5639)
represents an excellent local laboratory of a young starburst in an extreme metal-poor environment \citep[e.g., Fig.~\ref{f:f1}; ][]{SanchezAlmeida2013,Elmegreen2016,Lehmer2021}. The tadpole appearance of this dwarf galaxy arises from an extraordinary starburst region at one end of the galactic disk with an inclination of 84$^\circ$. 
21 young super stellar clusters of average mass $\sim 2 \times 10^4 {\rm~M_\odot}$ and age $\sim 5 \times 10^5$~yr are identified within the starburst region \citep[e.g., Fig.~\ref{f:f1}B-C][]{Elmegreen2016}. The line-of-sight visual extinction toward the stellar clusters is in the range of 0.6-0.8 mag.  Of particular interest is the region's notably low metallicity, 12 + log (O/H) $= 7.48 \pm 0.04$ \citep{SanchezAlmeida2013}, compared to the solar oxygen abundance, 12 + log(O/H)$_\odot = 8.69 \pm 0.05$ \citep{Asplund2009}, representing merely about 40\% of the corresponding value for the host galaxy \citep{SanchezAlmeida2013,Elmegreen2016}. Consequently, it is suggested that the starburst was instigated by a recent influx of low-metallicity gas, characterized by a total gas mass of $\sim 6 \times 10^6 {\rm~M_\odot}$. Although such an occurrence of localized cold accretion is infrequent in the present universe, it serves as the prevalent mechanism for gas supply in high-redshift galaxies \citep[e.g.,][]{SanchezAlmeida2013,Elmegreen2007}. These characteristics of \sou\ could be instrumental in refining the understanding of the HMXB population and offer valuable insights into galaxy formation and evolution in the early universe.

Furthermore, we noticed that an archival 7.9 ks \chandra/ACIS-S observation taken in 2018 showed the presence of a prominent pointlike source (hereafter called \xs) in the starburst head. With a 0.5-3 keV count rate of $24\pm 1.7 {\rm~cts~ks^{-1}}$, corresponding to $L_{\rm 0.5-10~keV} \sim 10^{40.3} {\rm~ergs~s^{-1}}$  (weakly depending on its spectrum) at a distance of \sou, this source is well within the range of the brightest 10\% ULXs or an extreme ULX. 

We have obtained new \chandra\ and Karl G. Jansky VLA observations of \sou\ to advance our understanding its properties.  This paper reports our study of \xs\ based on all \chandra\ data (Table~\ref{t:obs}), together with a complementary VLA observation. In the following, we describe the observations and data reduction in \S~2, present the data analysis and results in  \S~3, discuss their implications in \S~4, and finally summarize our main findings in \S~5. 

\section{Observations and Data Preparation}\label{s:obs}
Our X-ray study is based on four observations taken in 2024, plus one additional 7.9 ks observation in 2018, all with the Advanced CCD Imaging Spectrometer (ACIS-S; Table~\ref{t:obs}) in the very faint mode. We reprocess the data with the standard pipeline of the \chandra\ Interactive Analysis of Observations (CIAO; version 4.16 CALDB 4.11.2), including the light curve cleaning with the routine {\small LC\_CLEAN}. Resulting in a total of 95 ks exposure. 

\startlongtable
\begin{deluxetable*}{lccccccc}
\tablecaption{{\sl Chandra} Observation (OBS) Log}
\tablehead{\colhead{OBS} &
{\sl Chandra} & \colhead{R.A. (J2000)\tablenotemark{a}} & \colhead{Dec. (J2000)\tablenotemark{a}} & \colhead{Roll Angle} & \colhead{Exposure\tablenotemark{b}} & \colhead{OBS Start Date} & \colhead{End Date} \\
\colhead{\#} & \colhead{OBSID} & \colhead{(h~~m~~s)} & \colhead{($^\circ~~~^{\prime}~~~^{\prime\prime}$)} & \colhead{($^\circ$)} & \colhead{(s)} & \colhead{(yyyy-mm-dd)} & \colhead{(yyyy-mm-dd)}
}
\startdata
I & 20355    &11 41 07.857 & +32 25 39.93 &   148.2 &   7913 &2018-03-03 &2018-03-03  \\
II & 27107    &11 41 07.865 & +32 25 27.63 &    95.6 &  19812 &2024-01-21 &2024-01-21  \\
III & 26676    &11 41 07.940 & +32 25 49.26 &   226.1 &  12795 &2024-04-12 &2024-04-12  \\
IV & 29366    &11 41 08.098 & +32 25 50.03 &   217.0 &  16377 &2024-04-14 &2024-04-14  \\
V & 27108    &11 41 08.008 & +32 25 49.53 &   226.6 &  37704 &2024-04-28 &2024-04-29  \\
\enddata
\tablecomments{All observations were obtained using the ACIS-S detector. 
\tablenotemark{a} The coordinates correspond to the time-averaged location of the optical axis (the on-axis position). \tablenotemark{b}The exposure
represents the live time (corrected for dead time) of each observation.}
\label{t:obs}
\end{deluxetable*}

Since the pointings of the observations are all within 0\farcm3 of each other, we merge them for image-related analysis. \sou\ was observed on-axis and well covered by the back-illuminated S3 chip. However, we also include the adjacent front-illuminated S2 chip in the imaging analysis. We first align individual observations relative to the one with the longest exposure, and then visually compare the merged count image with \hst\ WFC3 images of the \sou\ field  \citep[e.g., Fig.~\ref{f:f1}A; ][]{Elmegreen2016} to find the discrete X-ray/optical counterparts for absolute astrometry correction. These images are Hubble Advanced Products \footnote{\url{https://mast.stsci.edu/search/ui/\#/hst}} and are aligned to a common astrometry reference frame, using Gaia as the reference catalog. We find two pairs of such counterparts, located outside \sou\ and at off-axis angles $\lesssim 2\farcm5$ of the \chandra\ observations. The PSF shape distortion at such small off-axis angles is negligible. The centroid positions of the X-ray sources are well determined. The offset between the X-ray and optical positions of the counterparts ($\approx 0\farcs2$, primarily to the north) are consistent. 

With the astrometry corrected data, we search for discrete X-ray sources in the three broad bands, 0.3-1.5 keV (S), 1.5-7 keV (H), and 0.3-7 keV (B), following the approach described in \citet{Wang2004}. This source detection reaches a 0.3-10~keV luminosity limit of $\sim 8 \times 10^{37} {\rm~erg~s^{-1}}$  (assuming the distance to \sou) with the local false detection probability of $10^{-7}$. 

\begin{figure}
\centerline{
\includegraphics[width=1.0\linewidth,angle=0]{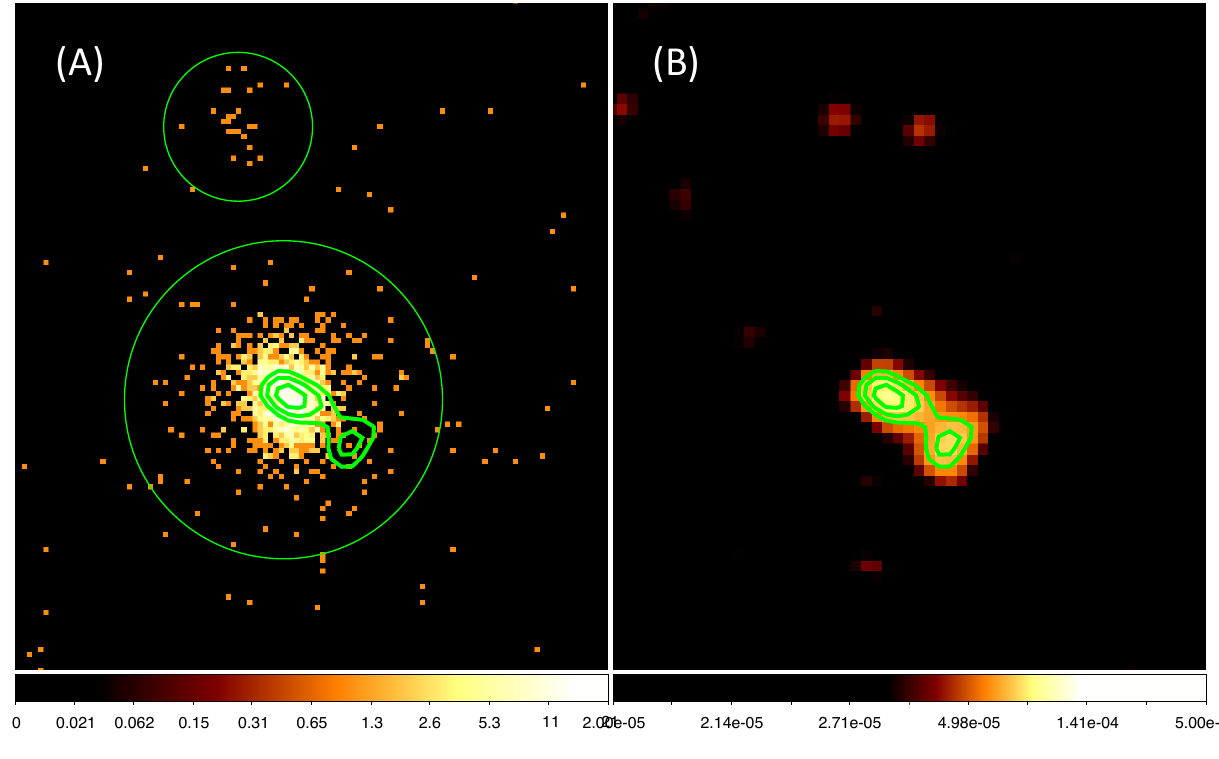}
}\caption{
{\bf (A)} \chandra\ ACIS-S 0.3-7~keV count image, which is constructed with all the observations listed in Table~\ref{t:obs}, together with the two detected discrete X-ray sources circled. The separation of them is 5\farcs2. {\bf (B)} VLA 1.5 GHz intensity image with the overlaid contours same as in Fig.~\ref{f:f1}B.}
\label{f:f2}
\end{figure}
Our \chandra\ observing program was accompanied by VLA observations to map the radio continuum and 21~cm line emission. The most relevant here is the A-array observation in the L-band (20\,cm; project code SO0060; observing date: 2023-07-04). This observation was set with 3C286 as the standard flux and bandpass calibrator and J1125+2610 as the complex gain/phase calibrator. The total observation time was 1\,hour, with about 30\,min at the source. The observation bandwidth is 1\,GHz in standard continuum mode. After flagging for radio interference and standard calibration, we obtain an image with $\sim 12\,\mu{\rm Jy~beam^{-1}}$, with a beam size of 1\farcs14 $\times$ 1\farcs08 and a position angle of $-0.61^{\circ}$ using a robust parameter of 0.5.

\section{Data analysis and Results}\label{s:res}

Fig.~\ref{f:f2} shows the \chandra\ ACIS-S 0.3-7~keV count image of the \sou\ field. X-1 is well detected at ${\rm R.A. = 11^h 41^m 07.^s52, DEC.} = +32^\circ 25^\prime 37\farcs4$ (J2000), coinciding with the \hst\ centroid of the central cluster within the position uncertainty $\sim 0\farcs15$  (90\% confidence; Fig.~\ref{f:f1}C). This source is also the brightest in the entire \chandra\ field. Another source can be seen at ${\rm R.A. = 11^h 41^m 07.^s59, DEC.} = +32^\circ 25^\prime 42\farcs6$ (uncertainty $\sim 0\farcm5$), or 5\farcs2 north of \xs\ in Fig.~\ref{f:f2},  is apparently outside the galaxy, and has no significant counterpart in the \hst\ WFC3 images. This source has a count rate a factor of $\sim 86$ lower than \xs\ or  $2 \times 10^{38}$\lum,  assuming the same count rate to the luminosity conversion inferred from the spectral modeling of \xs\ (see below).  Since the source is unlikely to be related to the starburst, we do not consider it further in this paper. 

\begin{figure}
\centerline{
\includegraphics[width=1.0\linewidth,angle=0]{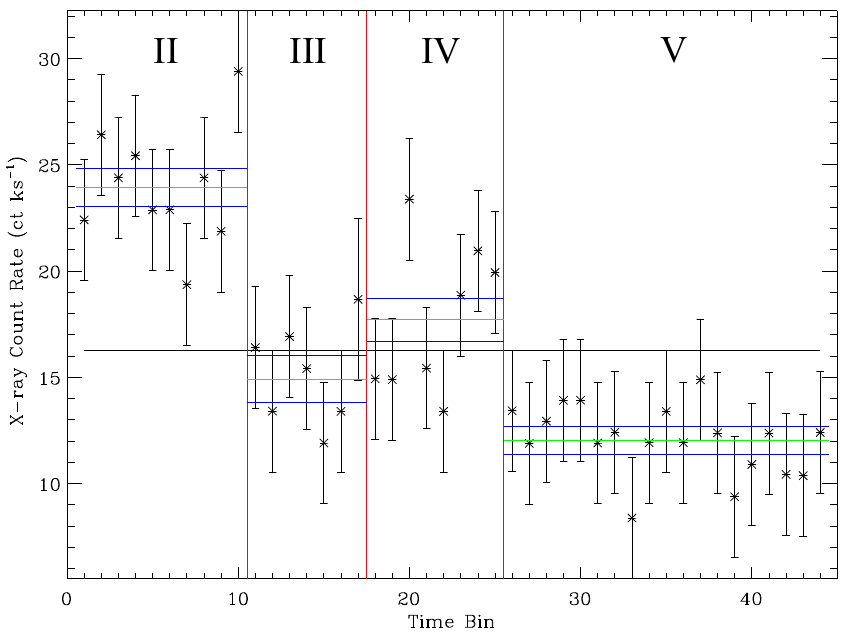}
}\caption{
Flux curve of \xs\ during OBS II-V, separated by the long vertical (red) lines. The count rates are calculated in the 0.3-7~keV band, and the local background is subtracted.  The long horizontal line marks the average count rate over the entire period of the four observations, while the short green and blue lines represent the average rate and its $1\sigma$ error bars of each observation.
}
\label{f:lc}
\end{figure}

We extract on-source counts of \xs\ from an aperture of the source removal radius ($\sim 2 \times$ the 90\% energy encircle radius of the \chandra/ACIS point spread function (PSF);  Fig.~\ref{f:f2}A) for our timing and spectral analyzes, The local background contribution is estimated in the surrounding region. Fig.~\ref{f:lc} presents the net flux curve of \xs, constructed from OBS II-V in the 0.3-7~keV band. The null hypothesis for a constant flux curve can be excluded at about 11$\sigma$ level ($\chi^2/d.o.f.=145/43$). The source varies significantly by up to a factor of $\sim 2$ between the observations, especially between OBS II and the remaining three. There is no evidence of significant variability within individual observations. Therefore, the variability is primarily on the time scale of months, corresponding to the time gaps between the observations. The variation also shows no evidence of any energy band dependence. We have excluded OBS I in this quantitative time analysis because it was taken about six years earlier than the other observations. Over this period, the effective area of the instrument changed substantially, which is energy dependent. With a crude broad-band correction of this effective area change, the flux of \xs\ during OBS I is consistent with the average value observed in OBS II-V (Fig.~\ref{f:lc}). Based on the detected variability and the dearth of similar point-like sources in the field, we conclude that \xs\ most likely represents a single ULX in the starburst region of \sou.

In addition, we find a source, 1eRASS J114107.4+322534, in the eROSITA All-Sky Survey (first 6 months, December 2019 - June 2020; eRASS1) X-ray (0.2-2.3 keV) source catalog \citep{Merloni2024} with a centroid offset $\sim 0\farcm19$ from \xs. The 0.2-2.3~keV count rate of this source is $0.203 \pm 0.053 {\rm~cts~s^{-1}}$, corresponding to a 0.3-10~keV flux that is about a factor of 2 higher than the mean flux observed during the \chandra\ observations. Considering the half-energy width $\sim 0\farcm5$ of the survey's PSF, the source should be dominated by \xs, although small contamination from a nearby diffuse X-ray feature $\sim 0\farcm5$ southwest of \sou\ is expected.  This is again consistent with the persistent nature of \xs, with only moderate variability.

For our spectral analysis, we use the software package \texttt{Xspec}  \citep[version 12.14.0; ][]{Arnaud1996}. 
We fit the local background subtracted spectrum of \xs\ with commonly used phenomenological spectral models that help to characterize the accretion state (Table~\ref{t:spec}). The foreground X-ray absorption is accounted for by the multiplicative model \texttt{TBabs} parameterized by the hydrogen column density ($N_H$), assuming the default solar metal abundances. Although a simple power law gives a statistically acceptable fit to the spectrum, a broken power law significantly improves the fit (Table~\ref{t:spec}). An F-test shows that the null hypothesis that the broken power law is not a statistically better fit than the power law has a probability of $8 \times 10^{-4}$.  An equally satisfactory fit to the spectrum is with the multicolor disk (MCD) model, \texttt{diskbb}, which is suitable for characterizing an accretion disk around a black hole accreting at a sub-Eddington rate \citep{Mitsuda1984,Makishima1986}. However, the fitted $N_H$ value is too small to be consistent with what  \citep[$\sim 10^{21} {\rm~cm^{-2}}$;][]{Bohlin1978} is expected from the visual extinction of 0.6-0.8 mag toward the starburst.
The slightly more flexible version of the MCD model \texttt{diskpbb} does not improve the fit statistically but allows $N_H$  to be more consistent with the extinction (Table~\ref{t:spec}). 
We further test the model, \texttt{diskbb+powerlaw}, and find that the power law component is largely negligible, affecting little the best-fit parameters of the disk component.  
In summary, the \texttt{diskbb}, \texttt{diskpbb}, and \texttt{bknpowerlaw} models give similarly satisfactory fits to the spectrum of \xs. Their estimates of the 0.3-10~keV luminosities ($\approx 10^{40.3} {\rm~erg~s^{-1}}$) are consistent with each other, insensitive to the exact value of $N_H$. We integrate the best fit \texttt{ diskpbb} model over the 0.1-50~keV, assuming $N_H = 10^{21} {\rm~cm^{-2}}$, to estimate the bolometric luminosity as $L_{bol}\approx 10^{40.4} {\rm~erg~s^{-1}}$. 
\begin{table*}
\caption{Primary spectral fit results of \xs\  \label{t:spec}}
\begin{tabular}{lcccr}
\hline\hline
Model & \texttt{TBabs*powerlaw}  & \texttt{TBabs*bknpowerlaw} & \texttt{TBabs*diskbb} & \texttt{TBabs*diskpbb}\\
\hline
Parameter & \multicolumn{4}{c}{Value}\\
$N_{\rm H}$ ($10^{21}$~cm$^{-2}$) & 3.2(1.9-4.6) & 0.25(0-2.1) & 0(0-0.45)& 0(0-1.5)\\
$\Gamma_1$ or $kT_{in}$ & 2.10(1.94-2.26) & 1.39(1.22-1.77) &1.3(1.2-1.4) &1.4(1.2-1.8)\\
$\Gamma_2$  & - &2.63(2.26-3.14) & -& -\\
Break E or $p$ & - & 3.0(2.6-3.6)& - & 0.68(0.57-0.80\\
$N (10^{-4})$\tablenotemark{a} &0.90(0.73-1.1)& 0.48(0.41-0.69)&43(35-55)&57(45-72)\\
$L_{0.3-10~{\rm keV}}$ & 40.57(40.48-40.61) & 40.35(40.32-40.44) &40.28(40.26-40.30)&40.30(40.27-40.39) \\
$\chi^2/d.o.f.$ & 142/132 & 127/130 &129/132&128/131\\
\hline
\\
\end{tabular}
\tablecomments{The parameter ranges included in the brackets are the 90\% confidence uncertainty intervals.
\tablenotemark{a} Normalization according to the definition of each \texttt{Xspec} model is not independent of the luminosity $L_{0.3-10~{\rm keV}}$ and is useful to constrain other interesting model parameters (e.g., see \S~\ref{ss:IMBH}).}
\end{table*}

\begin{figure*}
\centerline{
\includegraphics[width=1.0\linewidth,angle=0]{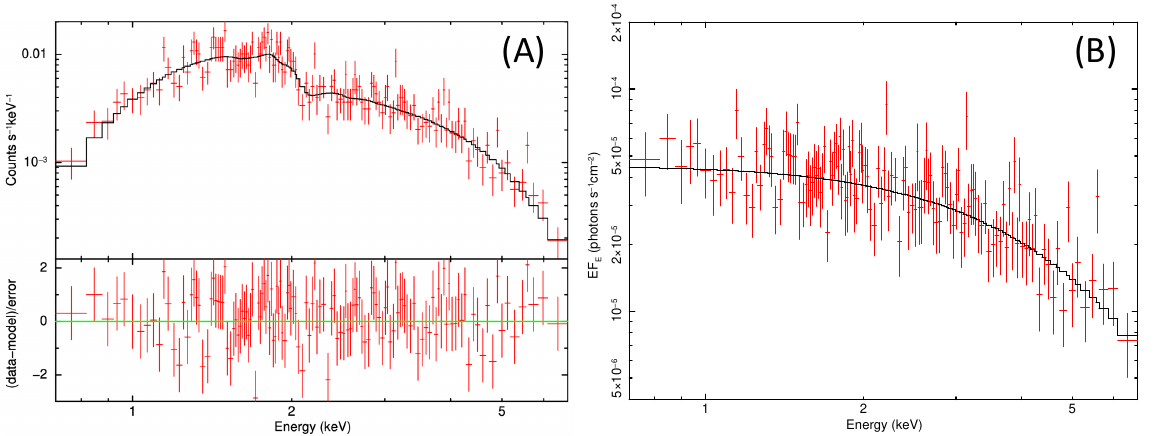}
}\caption{
{\bf (A)} \chandra\ ACIS-S spectral data of \xs, together with the best-fit MCD model \texttt{TBabs*diskpbb} (black line; Table~\ref{t:spec}); {\bf (B)} The "instrumental effect-free" best-fit model and the spectral data unfolded with the model, using the \texttt{Xspec} plot command 'eufspec'.
}
\label{f:spec-ULX}
\end{figure*}

The radio emission from the starburst region of \sou\ shows a double-lobed structure in the VLA L-band (effectively at 1.5\,GHz; Fig.~\ref{f:f2}). Accounting for nearly 2/3 of the total flux of the emission, the extended NE lobe is oriented in the NE-SW direction, and it positionally matches the centroid of \xs\ within its astrometry uncertainty (Fig.~\ref{f:f2}A).  The peak intensity of the lobe is 0.086~mJy. The SW lobe does not have any obvious counterpart in optical and may be powered by a jet from the source.  The separation between the two lobes is about 200~pc. We find that radio emission is also detected in the Low-Frequency Array (LOFAR) Two-metre Sky Survey (LoTSS), which covers the 120-168 MHz range \citep{Shimwell2017}. The intensity of the LoTSS peak of the emission is 0.69 mJy. The spatial resolution of the LoTSS (6\as) does not allow us to separate the two lobes. But assuming that they have the same spectral shape, we can estimate their relative contributions according to those in the 1.5 GHz image and hence the NE lobe intensity of 0.45~mJy in the LoTSS band.  We then estimate the spectral index $\alpha \approx 0.72$ of the emission (adapting the convention $S_\nu \propto \nu^{-\alpha}$), which is a typically observed value for active galactic nuclei in LOFAR surveys. The spectral index aligns with the average value of the NE lobe discerned in the VLA L-band index mapping. Nonetheless, the mapping reveals a trend of an increasing index with the increasing distance from the centroid of the NE lobe towards the SW, culminating at a certain value $\alpha \sim 3.7$. However, the restricted frequency range of the L-band hinders us from drawing a definitive conclusion.

\section{Discussion}\label{s:dis}


As an X-ray luminosity-selected source population, ULXs are rare in nearby galaxies and remain poorly understood \citep[e.g.,][]{Kaaret2017,King2023}.  They tend to be observed in actively star-forming galaxies, including a few dwarf galaxies \citep{Ott2005,Lehmer2022}, although they hardly have metallicities as low as \sou. The wide range of their X-ray spectral and variability characteristics suggests that they are a heterogeneous class \citep{Bachetti2016}. Their interpretations range from HMXBs, with stellar-mass compact objects accreting from their companions at super-Eddington rates and/or with anisotropic or relativistically beamed emission, to IMBHs \citep[e.g.,][]{Kaaret2017}.  In the following, we explore the nature of \xs\ based on the properties of this extreme ULX and its multi-wavelength counterparts. 

\subsection{Possibility of a ULX pulsar}\label{ss:ULXP}

A large fraction of ULXs are expected to be accreting neutron stars, although their pulsed emission is difficult to detect \citep[e.g.,][]{Misra2024}. Only several extragalactic ULXs have been identified to have pulsed X-ray emission that is clearly due to accreting neutron stars or pulsars \citep[e.g.,][]{Bachetti2014,RodriguezCastillo2020}. Some more candidates have been found through the tentative identification of pulsations \citep[e.g.,][]{Roberts2023}. These identified or candidate ULX pulsars (or ULXPs) are often extreme ULXs. Could \xs\ be such a ULXP?
 ULX population synthesis modeling shows that ULXPs are typically not important for a starburst with age $t \lesssim 10$~Myr, although the first ULXPs could form as early as 6~Myr  \citep{Wiktorowicz2017}. BHs dominate early epochs after a starburst, while ULXPs outweigh them after a few 100\,Myr and may appear as late as a few Gyrs after the end of the SF episode. The existing study of the starburst in \sou\ shows that it began only about $1 \times 10^7$~yrs ago and peaked at $t \sim 8 \times 10^5$~yrs \citep{Elmegreen2016}. The location of \xs\ is at the starburst's youngest and most massive central cluster (Fig.~\ref{f:f1}B). This consideration suggests that \xs\ is unlikely to be a ULXP. 

Another consideration is the variability behavior of \xs. A common characteristic of known ULXPs is their strong long-term variability, sometimes including transitions to 'off-states' in which their X-ray flux drops by orders of magnitude; the nature of such variations is not clear, which might be due to superorbital periodic, magnetic propeller, and/or obscuration effects \citep[e.g.,][]{Furst2023}. As shown above, the variability of \xs\ is present, but its flux change appears to be small (a factor of $\lesssim 2$) on time scales from months, although the sampling of the existing \chandra\ observations is still very limited. Considering this weak variability, as well as the youth of the starburst, the chance for \xs\ to be a ULXP is very small. 

\subsection{Possibility of a stellar-mass BH undergoing supercritical accretion}\label{ss:SBH}  
If \xs\ is an accreting stellar mass BH with $M \lesssim 10 M_{\odot}$ \citep[e.g.,][]{King2023}, its apparent extremity of high luminosity would mean that it is undergoing super-Eddington accretion, probably with the X-ray emission subject to a moderate degree of geometric beaming \citep{King2001,Begelman2002,Poutanen2007}. Statistically, it has been shown that more extreme beaming from relativistic jets is unlikely \citep{King2004}. 
The super-Eddington accretion, when advection suppresses the emission of the innermost disk regions, may be modeled by a slim disk \citep[e.g.,][]{Abramowicz1988}. We use the results of the spectral fits with the \texttt{diskpbb} model  (Table~\ref{t:spec}) to distinguish the potential states of the accretion disk \citep[e.g.,][]{Yoshimoto2024}. This spectral model fits the \chandra\ spectrum of \xs\ well. The fitted $p = 0.68 (0.57-0.80)$ is consistent with the prediction of the standard optically thick and geometrically thin disk \citep{Shakura1973} and is inconsistent with that of the optically thin and geometrically thick (slim) disk  ($p = 0.50$)   \citep{Watarai1999,Watarai2000}.  However, the fit does not show evidence for a soft excess or a high-energy break, or significant deviation from the \texttt{diskpbb} model fit, as shown in the modeling of high-quality spectral data of ULXs that are likely due to super-Eddington accretion \citep{Gladstone2009}. So, considering such uncertainties in the existing data and theories \citep[e.g.,][]{Ebisawa2003}, we cannot firmly rule out the possibility that \xs\ is a stellar-mass BH undergoing a super-critical accretion. 

\subsection{Possibility of an IMBH with a standard accretion disk}\label{ss:IMBH}

The high luminosity of \xs\ and the consistency of its X-ray spectrum with the standard accretion disk model suggest that it could be an IMBH undergoing sub-Eddington accretion. 
We deduce the properties of this putative IMBH. 
 Assuming our estimated bolometric luminosity and that the accretion is close to the Eddington limit of the IMBH, we constrain its minimum mass to be $\gtrsim 1.7 \times 10^2 {\rm~M_\odot}$. This mass can then be used to estimate the radius of the innermost stable circular orbit of the accretion disk as $R_{in} \gtrsim 1.5 \times 10^3$~km if the IMBH has no spin or up to a factor of 6 smaller (i.e., $R_{in} \gtrsim 2.6 \times 10^2$\,km) if it has a maximum spin.
$R_{in}$ can also be constrained by our spectral fit parameter $N$ (Table~\ref{t:spec}), since $R_{in}=\xi \kappa^2 r_{in}$, while $r_{in}= D_{10}(N/{\rm cos}i)^{1/2}$, where $i$ is the inclination angle of the disk and $D_{10}$ (in units of 10~kpc) is the distance of \xs, here assumed to be the same as that of \sou. We adopt the boundary correction factor $\xi =0.41$ \citep{Kubota1998}, the hardening factor $\kappa = 1.7$ for the color temperature correction \citep{Shimura1995}. The best fit \texttt{diskpbb} model gives $N \approx 5.7 \times 10^{-3}$, 
when our preferred prior values $N_H=10^{21} {\rm~cm^{-2}}$ and $p=0.75$ are assumed. With these parameters, we obtain $R_{in}= 2.2\times 10^2({\rm cos}i)^{-1/2}$~km, which is consistent with the value inferred above from the luminosity and mass of the IMBH, if $i \gtrsim 39^\circ$ or $89^\circ$ for maximum or no spin.
Fast spin allows for a hot accretion disk with a small $R_{in}$ \citep{Makishima2000}, which explains the atypically high $T_{in}\sim 1.3$ of \xs. Therefore, \xs\ could represent a fast-spinning IMBH, accreting sub-critically via a reasonably highly inclined disk.

\subsection{Implications of the radio counterpart for \xs}
Assuming that the detected non-thermal radio emission is physically associated with \xs, we can independently constrain the mass of the accreting BH ($M_{BH}$). We first estimate the 5 GHz luminosity as $L_R = 1.3 \times 10^{35}$\lum, based on the NE lobe intensity of the emission at 1.5 GHz and our estimated $\alpha = 0.72$.  Empirically, this luminosity could only be produced by jets from a sub-Eddington-accreting IMBH or supermassive BH (SMBH) \citep[e.g.,][]{Panurach2024}.  We then estimate the 2-10~keV luminosity as $L_X = 1.2 \times 10^{40}$\lum, based on the best fitting \texttt{diskpbb} model (Table~\ref{t:spec}). Using the $M_{BH}-L_X-L_R$ fundamental plane for BH activity \citep[][their Eq. (8) or (15), for their entire source sample or Seyferts alone, respectively]{Gultekin2019}, we estimate $M_{BH} = 10^{5.4} {\rm~or~} 10^{5.2} M_\odot$. This estimate is insensitive to the exact $\alpha$ value adopted; changing it by $\pm 0.3$ would change the mass estimate by 0.17 dex, for example. The relationship's dominant uncertainty is the intrinsic scatter of the data (close to 1 dex). Therefore, the mass estimate is consistent with the hypothesis that \xs\ is an accreting IMBH or SMBH.  AGNs with SMBHs have been observed in dwarf galaxies \citep[e.g.,][]{Mezcua2023} and some even appear in regions far from galactic centers in nearby galaxies \citep[e.g.,][]{Secrest2017}. However, the association of such an SMBH with an extremely young, low metallicity, off-center starburst, as seen in \sou, would be unique to our knowledge and probably difficult to explain.  Therefore, our preferred scenario for \xs\ is an accreting IMBH undergoing sub-Eddington accretion.

The extended radio emission also excludes alternative scenarios that may be proposed for \xs. One hypothesis could be that \xs\ is a nascent supernova remnant (SNR) with an age of $\lesssim 10^2$ years. Given the intense shock interaction between the supernova ejecta and the dense circumstellar medium, such an SNR could potentially attain a luminosity of $10^{39} {\rm~erg~s^{-1}}$ or greater, despite the absence of direct observational corroboration to the best of our knowledge. The \chandra\ spectrum of \xs\ can be well modeled using an optically-thin thermal plasma model \texttt{apec} with a temperature of $4.3$ keV and a metal abundance of $\lesssim 0.2$ solar, characteristics that align with this scenario. However, the luminosity of the SNR is expected to initially escalate rapidly and then gradually diminish over several years, contradicting the observed long-term persistency and variability over a monthly timescale. Furthermore, a young SNR would typically manifest as a point-like radio source, contrary to the observed extended and elongated structure. Considering these factors and the observed extreme X-ray luminosity, we conclude that \xs\ is highly unlikely to be a young SNR.

\subsection{Formation of \xs}

The existing age estimate of the central cluster of the starburst in \sou\ is far too small to allow the formation of the BH (let alone a neutron star) from normal stellar evolution even for the most massive stars. The lifetime of a $100~M_\odot$ star, for example, is about 2.5~Myr \citep[e.g.,][]{Leitherer1999}. Constraints on the age of the starburst can also be obtained from the existing Sloan Digital Sky Survey (SDSS) spectrum (with a fiber diameter of 3\as covering most of the starburst), which shows prominent emission lines He\texttt{II}$\lambda$4658 and [Ar\texttt{IV}]$\lambda$4711 \citep{SanchezAlmeida2013}, indicating the strong presence of ionization radiation at energies $> 54.4$ eV and 59.8 eV. Interestingly, the spectrum does not indicate the broad features expected for Wolf-Rayet stars, which would be an important source of extreme ultraviolet (EUV) radiation. This strong presence of highly ionized species without evident Wolf-Rayet features is not unusual in low-metallicity dwarf galaxies \citep{Shirazi2012}. In the starburst head of \sou, much of the ionization radiation could arise from \xs. 

We suspect that the existing age estimate of the clusters in the starburst head, which is based on the color-magnitude analysis of the cluster \citep{Elmegreen2016}, might be strongly affected by the presence of ULX \xs, especially its UV emission \citep[e.g.,][]{Sonbas2019}, and/or the dynamical effect on the stellar properties \citep[e.g., due to runaway mergers and stripping stellar envelopes;][]{Vanbeveren2009,Gotberg2018}.  A more probable age of the starburst is $\sim$ 3 Myr with an uncertainty of about 1 Myr, consistent with the accreting object of \xs\ as the end product of a very massive star and with the lack of the Wolf-Rayet signature, which is expected to occur at a cluster age of 4-7 Myr for such a low-metallicity cluster as in the starburst head of \sou.  At an older age,  ionizing radiation is expected to have decreased substantially, inconsistent with the strong H$\alpha$ emission of the starburst.  The moderate age of the starburst is also required to explain our estimated mass accretion rate of \xs, $\dot{M} \approx (6.7$ or $1.1) \times 10^{-6} {\rm~M_\odot~yr^{-1}}$, where we have assumed a BH accretion with our inferred bolometric luminosity of \xs\ ($L_{bol}\approx 10^{40.4} {\rm erg~s^{-1}}$; \S~\ref{s:res}) and the energy conversion efficiency of optically thick accretion disks as $\eta =$ 0.0625 or 0.366 for the Schwarzschild or extreme Kerr case, respectively \citep{Ebisawa2003}.  The estimated rate strongly indicates a Roche lobe filling accretion from an evolved star, a case that is probably similar to that for ULX Ho II X-1 \citep{Serantes2024}.

ULXs are believed to be formed largely via dynamical processes in stellar clusters. Statistical studies \citep[e.g.,][]{Poutanen2013} have shown a highly significant association between ULXs and young clusters. Their typically large displacements (e.g., up to 300\,pc) indicate that the most massive stars (binaries) are ejected at the beginning stages of the cluster evolution due to close encounters. In particular, HLXs have been suggested to be potentially consistent with formation within globular clusters \citep[e.g.,][]{Barrows2024}. What can we learn from the presence of \xs\ in the young starburst head of \sou? 

The formation of ULX \xs\ in \sou\ is likely related to its recent accretion of low-metallicity gas \citep[e.g.,][]{Nakajima2024}. This gas may originate in a very outer region of a neighboring galaxy, e.g., via tidal stripping and accretion. Indeed, our 21~cm survey shows the presence of neighboring dwarf galaxies with evidence for tidal interactions. Although the study of these interactions is still ongoing, we speculate that the starburst in \sou\ is triggered by such a recent cold gas accretion event. Previous X-ray observations have already shown indications of the increased probability of finding ULXs in metal-poor galaxies \citep[e.g.,][]{Prestwich2013,Ponnada2020,Cann2024}.  Furthermore, simulations have shown that supercompetitive accretion to central supermassive stars can occur in a massive dense cloud as long as its metallicity is low enough ($\lesssim 10^{-3} Z_\odot$), which may even lead to the formation of IMBHs \citep[e.g.,][]{Alexander2014,Chon2020,Latif2022}. The presence of IMBHs remains a hotly debated topic, although their existence in some nearby dwarf galaxies has been proposed \citep[e.g.,][]{Greene2020,Eberhard2024}. Strong observational evidence for their existence in regions off galactic nuclei remains particularly rare, probably except in the stellar cluster $\Omega$ Centauri in our Galaxy \citep{Haberle2024}. The deep, high-resolution X-ray study of \sou\ presented here provides at least an empirical insight into what might be happening in more distant dwarf galaxies with similar X-ray enhancements \citep[e.g.,][]{Cann2024}.  

While the formation of ULXs in extremely young clusters remains poorly understood \citep[e.g.,][]{Sgalletta2024}, developing a model specifically for the formation of \xs\ is certainly beyond the scope of the present work. Nevertheless, we note that our age estimate up to $\sim 4$~Myr for the starburst may make the dynamical formation of an IMBH feasible, depending on the initial conditions and environment of the host stellar cluster. Its low metallicity also helps to retain more massive stars, increasing encounter rates and enhancing the dynamical process. All of these effects may have contributed to the formation of \xs. In the process, the surrounding gas could have been significantly enriched by the massive stars via stellar winds and potentially pair-instability supernovae, which could occur for stars of masses in the range of $\sim 130-250 {\rm~M_\odot}$ and extremely low metallicities \citep[e.g., $\lesssim 10^{-2} - 10^{-3}$ solar; ][]{Heger2003,Gabrielli2024}.

\subsection{Future prospects}

More observations are clearly needed to improve our understanding of ULX \xs\ in \sou. Our existing detection of variability on time scales of months is consistent with those observed in ULXs with similar luminosities (e.g. I Zw18 ULX) \citep[e.g.,][]{Yoshimoto2024}. But the counting statistics of the existing \chandra\ observations of \xs\ are too limited to constrain on its spectral variation tightly. X-ray observations with improved counting statistics will further enable the study of the luminosity and spectral variation on timescales from days to years. The bolometric luminosity follows $L_{bol} \propto T^4$ for a standard disk \citep{Shakura1973}, whereas $L_{bol} \propto T^2$ is expected for a slim disk, where advection decreases the radiation efficiency \citep{Abramowicz1988,Watarai2000}. The strongest evidence for supercritical accretion would be the detection of emission and absorption lines originating in a fast outflow, which are highly redshifted or blueshifted, as observed in several ULXs \citep[e.g.,][]{Pinto2016}. 
Most exciting would be the confirmation of the accreting IMBH nature of \xs\  -- our favored scenario. 
The detection of quasi-periodic oscillations in \xs\ would help us to distinguish an IMBH (e.g., in the mHz range) from a stellar-mass one (at higher frequencies) \citep[e.g.,][]{Gladstone2009}. A similar time scale-based distinction can also be made in detecting the state transition or characteristic break frequency in the power density spectrum of the X-ray variability. Ideally, the IMBH nature could be confirmed dynamically via observations of stellar and accretion gas kinematics of \xs.

Although our existing data and analysis disfavor the ULXP scenario, a search for the periodic signal and long-term monitoring of \xs\ is desired for further tests. The formation of neutron stars (and hence the presence of a ULXP) in the starburst head cannot be completely ruled out, especially since the lifetime of an extremely low-metallicity star could be significantly shortened and since the age estimation of the star clusters could be uncertain. The pulse periods of known ULXP range from 0.4 to 40\,s, while the pulsed fractions are $f_{p} \approx 10\%$ \citep[e.g.,][]{RodriguezCastillo2020}.  A deep {\sl XMM-Newton} pn observation will allow a sensitive search for the pulsed signal.  If the pulsation is detected, it will {\it prove} that \xs\ is an accreting pulsar produced in a young low-metallicity starburst and will be a very interesting addition to the existing very limited sample of known ULXPs.  Furthermore, for stellar-mass systems, orbital modulations in the X-ray flux curve may be detectable on timescales of hours to days. 

Multi-wavelength observations of \sou\ can be exceptionally beneficial. To improve our understanding of the structural and spectral characteristics of the radio counterpart of \xs, it is paramount to acquire more refined data, such as VLA C-band observations in A and B configurations. Conducting spatially resolved optical or ultraviolet (UV) spectroscopy will further contribute to accurately determining the stellar and interstellar attributes of the starburst. Elucidating the source of EUV radiation in this region and analogous starbursts within nearby dwarf galaxies will offer insights into phenomena occurring in high-z galaxies \citep[e.g.,][]{Garofali2024,Bi2024}. The strong presence of high-ionization nebular emission lines in such galaxies has been observed over the past decade, especially with the advent of \jwst, and remains poorly understood. At present, there is no consensus on the importance of ULX in explaining He$^+$ line emission (e.g., at $\lambda =4686$ \AA\ due to the transition $n=4-3$), for example, \citep[e.g.,][]{Simmonds2021,Gurpide2024,Garofali2024}. ULXs show a wide range of spectral energy distributions (SEDs) and EUV luminosities. The SED modeling of distant galaxies depends on the knowledge of the underlying stellar binary evolution, the X-ray source population, and the gaseous environment, which can hardly be well constrained, both theoretically and observationally, especially for distant galaxies \citep[e.g.,][]{Stanway2019,Tsai2023,Garofali2024}. There are also additional sources of EUV emission arising from fast interstellar shocks and diffuse hot plasma, which may be enhanced in low-metallicity starbursts \citep[e.g.,][]{Oskinova2022}. In short, spatially resolved optical and/or UV spectroscopy of nearby galaxies, ideally studied statistically, is desirable to improve our understanding of the origins of the high-ionization ions and ULXs, as well as their relationship to the metallicity in starbursts. 
\section{Summary}\label{s:sum}

We have obtained 90\,ks \chandra\ data, plus a VLA A-array 1.5~GHz image, on \sou\ to conduct an in-depth study of the high-energy end-products of its starburst region with about 6\% solar metallicity. Our main results and conclusions in this work are as follows.
\begin{itemize}
    \item A dominant point-like source, \xs, is detected with a mean luminosity of $\sim 10^{40.3}$\lum\ in the 0.3-10 keV band. This source exhibits significant variability over months, with an intensity change of a factor $\sim 2$.
    \item The X-ray spectrum of \xs\ is well-modeled by a multi-color blackbody accretion disk. The spectrum and extreme luminosity are most consistently explained by a fast-spinning IMBH with a sub-Eddington accretion disk viewed at a reasonably high inclination angle. However, a scenario of strongly beamed emission from a stellar-mass compact object (neutron star or BH) cannot be ruled out. 
    \item X-1 also has an apparent radio counterpart, which shows a double-lobed morphology extending $\sim 200$ pc and a nonthermal spectral index $\alpha \approx 0.7$, indicating a steepening with increasing distance from the source.  This radio counterpart,  probably the longest known for a ULX, may represent a powerful jet. The combination of our measured radio and X-ray luminosities, as well as the $M_{BH}-L_X-L_R$  fundamental plane for BH activity, gives an estimate of the BH mass of a few $\times 10^5 M_\odot$. With an uncertainty of about one dex, this estimate is consistent with the IMBH hypothesis of \xs.
    \item X-1 is found to coincide positionally with the central cluster of the starburst with an existing age estimate of only $8 \times 10^5$~year. This age may be a substantial underestimation, presumably because of enhanced UV emission that has not been accounted for. This enhancement is evidenced by the prominent presence of highly ionized species (He\texttt{II}$\lambda$4658 and [Ar\texttt{IV}]$\lambda$4711) observed in an SDSS spectrum of the starburst. It can potentially arise from the ULX and/or stellar dynamics of the cluster. The spectrum shows no broad features expected for Wolf-Rayet stars, which could otherwise be an important source of extreme UV (EUV) radiation. We estimate the age of the starburst to be about 2-4 Myr to allow for the minimum time for massive stars to evolve into \xs\ and to account for the lack of the Wolf-Rayet features. 
    \item The starburst in \sou\ is most likely induced by a major recent influx of low-metallicity gas, which may be part of a tidally disrupted outer disk of a neighboring dwarf galaxy. The initial metallicity of this gas could be extremely low ($\lesssim 10^{-2} - 10^{-3}$ solar), leading to the formation of the IMBH in the starburst. The gas could have subsequently been enriched by evolved massive stars via stellar winds and potentially supernovae.
\end{itemize}
Although further studies are needed to determine the exact nature of \xs, this work demonstrates that the starburst in \sou\ provides an ideal setting to reveal end products, especially the possible formation of an IMBH, in a young low-metallicity starburst. A more thorough analysis of \sou, including diffuse X-ray emission in and around it, is ongoing and will be presented later. The radio continuum and 21\,cm line mapping will complement this analysis by combining the VLA A, B, C, and/or D configuration data. These studies will offer insight into high-energy astrophysical phenomena and processes and their interplay with galactic environments in high-z galaxies.

\section*{Acknowledgements}
We thank the anonymous referee for constructive comments that improved the manuscript. QDW appreciates the discussion with Ken Ebisawa, Ken Chen, and Sunmyon Chon and the support provided by the Academia Sinica Institute of Astronomy and Astrophysics, where part of this work is completed. 
We acknowledge the support for this work provided by the National Aeronautics and Space Administration via the grant G03-24062X through the Chandra X-ray Observatory Center, which is
operated by the Smithsonian Astrophysical Observatory for and on behalf of the National Aeronautics and Space Administration under contract NAS8-03060, and by the National Radio Astronomy Observatory, which is a facility of the National Science Foundation operated under cooperative agreement by Associated Universities, Inc.

\section*{Data Availability}
This paper employs a list of Chandra datasets, obtained by the Chandra X-ray Observatory, contained in~\dataset[DOI: 10.25574/cdc.340]{https://doi.org/10.25574/cdc.340}.

%

\vspace{5mm}
\facilities{Chandra and VLA
}


       
\bibliography{ms-v1-astro-ph.bib}{}
\bibliographystyle{aasjournal}

\end{document}